\documentstyle[12pt]{article}

\newcommand{\al}{\alpha}
\newcommand{\be}{\beta}
\newcommand{\si}{\sigma}

\newcommand{\eqn}[1]{(\ref{#1})}  
\newcommand{\Eqn}[1]{Eq.~(\ref{#1})}  

\newcommand{\beq}{\begin{equation}}
\newcommand{\eeq}{\end{equation}}
\newcommand{\ba}{\begin{array}}
\newcommand{\bea}{\begin{eqnarray}}
\newcommand{\ea}{\end{array}}
\newcommand{\eea}{\end{eqnarray}}

\newcommand{\comment}[1]{ \hbox{[{\it Comment suppressed here.}\/]} }
\newcommand{\hide}[1]{}

\newcommand{\Tr}{\hbox{Tr}}

\renewcommand{\Re}{ {\rm Re}\, }
\newcommand{\bearray}{\begin{eqnarray}}
\newcommand{\eearray}{\end{eqnarray}}
\newcommand{\nl}{\nonumber \\}
\newcommand{\pl}{{\rm pl}}
\newcommand{\rt}{{\rm rt}}
\newcommand{\trt}{{\rm trt}}
\newcommand{\pg}{{\rm pg}}
\newcommand{\ls}[1]{\mbox{$\frac{1}{3}$\,Re\,Tr}(1-#1)}
\newcommand{\order}{{\cal O}}
\newcommand{\third}{{\textstyle {1\over 3}}}
\newcommand{\dgsp}{\mbox{$\phantom{0}$}}

\begin{document}

\title{Lattice QCD on Small Computers}

\author{M.~Alford, W.~Dimm and G.P.~Lepage \\
 \small Floyd R. Newman Laboratory of Nuclear Studies \\
  \small Cornell University, Ithaca, NY 14853 \\[2ex]
 G.~Hockney and P.B.~Mackenzie \\
 \small Theoretical Physics Department \\  \small Fermilab,
  Batavia, IL 60510
\\[2ex]
}

\newcommand{\preprintno}{
to be published in Physics Letters B \\[0.5ex]
\small FERMILAB-Pub 95/199-T \\[0.5ex] {\tt hep-lat/9507010}
}

\date{\small August 1995 \\[1ex] \preprintno}

\begin{titlepage}
\maketitle

\begin{abstract}
We demonstrate that lattice QCD calculations can be made $10^3$--$10^6$
times faster by using very coarse lattices. To obtain accurate
results, we replace the standard lattice actions by
perturbatively-improved actions with tadpole-improved correction
terms that remove the leading errors due to the lattice.
To illustrate the power of this approach, we calculate the
static-quark potential,
and the charmonium spectrum
and wavefunctions using a desktop computer. We obtain accurate
results that are independent of the lattice spacing and
agree well with experiment.

\end{abstract}

\end{titlepage}

\section{Introduction}

For 20 years, quantum chromodynamics (QCD) has been the generally
accepted theory of strong interactions in particle physics. However,
physicists are only just beginning to show that it explains basic features of
low-energy strong-interaction physics, such as the spectrum and structure of
the hadrons. The most significant progress in the study of low-energy
(nonperturbative) QCD has come from numerical simulations of a lattice
approximation to the theory. However this approach has been severely
limited by the rapid rise in computational difficulty as one
approaches the physically relevant limit of small lattice spacing and
large lattice volume. In particular the cost of lattice simulations
typically grows as $a^{-6}$ when the lattice spacing~$a$ is reduced.
With standard lattice techniques, it is widely felt that lattice
spacings of 0.05--0.1~fm or less are necessary for reasonable accuracy,
making simulations impossible on all but the largest computers.
In this paper, we suggest modifications of lattice QCD that permit
accurate simulations at lattice spacings as large as 0.4~fm,
increasing the speed of simulations by a factor of $10^3$--$10^6$.
We illustrate our approach with state-of-the-art calculations of the
static-quark potential and the charmonium spectrum performed on a
high-performance personal computer\,\cite{gplbeil}.

The central approximation in lattice QCD is replacing continuous
spacetime by a discrete lattice. The QCD action is discretized by
replacing space-time
integrals with sums and derivatives with differences.
Then the path integral defining the field theory can be evaluated
numerically using Monte Carlo techniques. Standard discretizations of
the QCD action have errors of $\order(a^2)$ that are large when
the lattice spacing is 0.4~fm.
However improved discretizations can be designed in which
finite-$a$ errors are
systematically removed by introducing new (nonrenormalizable)
interactions into the lattice action.
This does not add new parameters to the theory, since the coefficients
of the new interactions are determined, from first principles,
by demanding that the discretized action reproduces continuum physics to a
given accuracy.
Since the new interaction terms correct for
deficiencies in the short-distance behavior of the lattice theory,
their coefficients can be computed using perturbation theory in
asymptotically free theories such as QCD, provided the lattice spacing
is small enough that perturbation theory is applicable at distances of
order~$a$ and smaller.

The use of perturbatively improved actions for lattice QCD was
suggested long ago\,\cite{sym}.
However early tests showed little benefit from the
improvements. Furthermore a variety of studies seemed to indicate that
perturbative dynamics was only relevant at distances significantly
smaller than 0.1~fm,  suggesting that lattice spacings would have to
be smaller as well. Recent work\,\cite{lm} on lattice perturbation theory,
however, has completely changed this conclusion.
If conventional lattice perturbation theory (in terms of the bare
coupling) is replaced by an expansion in terms of a renormalized
coupling, perturbation theory becomes useful even at distances as
large as 0.5~fm. If in addition lattice operators are ``tadpole
improved'' the convergence of the perturbative expansions needed to
define improved actions is greatly enhanced.
Indeed, as we show here,
tree-level calculations of the improvements suffice in most cases provided
operators are tadpole-improved. Without tadpole improvement, the
correction terms are systematically underestimated, sometimes by
factors of two or three.

Tadpole improvement is a technique for summing to all orders the
large perturbative contributions that arise from tadpole diagrams
peculiar to lattice QCD. At tree level the improvement is trivial to
implement: one replaces each lattice QCD link operator $U_\mu$
in the action by $\tilde U_\mu=U_\mu/u_0$, where $u_0$ is
a scalar mean value of the link, defined to be the fourth root of
the expectation value of the four-link plaquette in Monte-Carlo
simulations\,\cite{lm}.

In several earlier papers,
we have advocated the use of perturbatively improved actions with
tadpole improvement\,\cite{gplbeil,lm,gpltasi,gpltsukuba}.
Such actions have already proven very successful in
simulations of heavy-quark mesons like the~$\Upsilon$. In the
nonrelativistic NRQCD quark action\,\cite{nrqcd}, both relativistic effects and
finite-$a$ corrections are introduced through nonrenormalizable
corrections to the basic action. The detailed simulation results
presented in~\cite{nrqcd} agree well with experiment, and many depend
crucially on these corrections.  Tadpole improvement was essential to
this success; test simulations without tadpole improvement underestimated
relativistic effects by as much as a factor of two. Similar results
have been obtained with improved versions of the Wilson quark action
when applied to heavy-quark mesons\,\cite{fermi}.

In this paper we present new evidence that
perturbatively-improved actions, once they are tadpole improved, work well
for gluons as well, even at spacings as large as $a=
0.4$~fm.
Using improved actions for the quarks and
gluons on a lattice with $a= 0.4$~fm, we obtain a static
potential that is rotationally invariant to within a few percent,
the spin-averaged charmonium spectrum accurate to within 30--40~MeV,
and rotationally invariant charmonium wavefunctions.
Our results show lattice-spacing independence (scaling) to within 3\%.

\section{The Improved Action}

\begin{figure}
\begin{center}
\setlength{\unitlength}{0.240900pt}
\ifx\plotpoint\undefined\newsavebox{\plotpoint}\fi
\sbox{\plotpoint}{\rule[-0.175pt]{0.350pt}{0.350pt}}%
\begin{picture}(1200,900)(0,0)
\tenrm
\sbox{\plotpoint}{\rule[-0.175pt]{0.350pt}{0.350pt}}%
\put(264,158){\rule[-0.175pt]{210.065pt}{0.350pt}}
\put(264,158){\rule[-0.175pt]{0.350pt}{151.526pt}}
\put(264,284){\rule[-0.175pt]{4.818pt}{0.350pt}}
\put(242,284){\makebox(0,0)[r]{$1$}}
\put(1116,284){\rule[-0.175pt]{4.818pt}{0.350pt}}
\put(264,410){\rule[-0.175pt]{4.818pt}{0.350pt}}
\put(242,410){\makebox(0,0)[r]{$2$}}
\put(1116,410){\rule[-0.175pt]{4.818pt}{0.350pt}}
\put(264,535){\rule[-0.175pt]{4.818pt}{0.350pt}}
\put(242,535){\makebox(0,0)[r]{$3$}}
\put(1116,535){\rule[-0.175pt]{4.818pt}{0.350pt}}
\put(264,661){\rule[-0.175pt]{4.818pt}{0.350pt}}
\put(242,661){\makebox(0,0)[r]{$4$}}
\put(1116,661){\rule[-0.175pt]{4.818pt}{0.350pt}}
\put(438,158){\rule[-0.175pt]{0.350pt}{4.818pt}}
\put(438,113){\makebox(0,0){$1$}}
\put(438,767){\rule[-0.175pt]{0.350pt}{4.818pt}}
\put(613,158){\rule[-0.175pt]{0.350pt}{4.818pt}}
\put(613,113){\makebox(0,0){$2$}}
\put(613,767){\rule[-0.175pt]{0.350pt}{4.818pt}}
\put(787,158){\rule[-0.175pt]{0.350pt}{4.818pt}}
\put(787,113){\makebox(0,0){$3$}}
\put(787,767){\rule[-0.175pt]{0.350pt}{4.818pt}}
\put(962,158){\rule[-0.175pt]{0.350pt}{4.818pt}}
\put(962,113){\makebox(0,0){$4$}}
\put(962,767){\rule[-0.175pt]{0.350pt}{4.818pt}}
\put(264,158){\rule[-0.175pt]{210.065pt}{0.350pt}}
\put(1136,158){\rule[-0.175pt]{0.350pt}{151.526pt}}
\put(264,787){\rule[-0.175pt]{210.065pt}{0.350pt}}
\put(45,472){\makebox(0,0)[l]{\shortstack{$a\,V(r)$}}}
\put(700,68){\makebox(0,0){$r/a$}}
\put(395,693){\makebox(0,0)[l]{a) Wilson Action}}
\put(264,158){\rule[-0.175pt]{0.350pt}{151.526pt}}
\put(438,293){\circle{12}}
\put(511,392){\circle{12}}
\put(566,478){\circle{12}}
\put(613,427){\circle{12}}
\put(654,510){\circle{12}}
\put(787,553){\circle{12}}
\put(962,636){\circle{12}}
\put(438,293){\usebox{\plotpoint}}
\put(428,293){\rule[-0.175pt]{4.818pt}{0.350pt}}
\put(428,293){\rule[-0.175pt]{4.818pt}{0.350pt}}
\put(511,388){\rule[-0.175pt]{0.350pt}{1.927pt}}
\put(501,388){\rule[-0.175pt]{4.818pt}{0.350pt}}
\put(501,396){\rule[-0.175pt]{4.818pt}{0.350pt}}
\put(566,473){\rule[-0.175pt]{0.350pt}{2.409pt}}
\put(556,473){\rule[-0.175pt]{4.818pt}{0.350pt}}
\put(556,483){\rule[-0.175pt]{4.818pt}{0.350pt}}
\put(613,426){\rule[-0.175pt]{0.350pt}{0.482pt}}
\put(603,426){\rule[-0.175pt]{4.818pt}{0.350pt}}
\put(603,428){\rule[-0.175pt]{4.818pt}{0.350pt}}
\put(654,508){\rule[-0.175pt]{0.350pt}{1.204pt}}
\put(644,508){\rule[-0.175pt]{4.818pt}{0.350pt}}
\put(644,513){\rule[-0.175pt]{4.818pt}{0.350pt}}
\put(787,545){\rule[-0.175pt]{0.350pt}{3.854pt}}
\put(777,545){\rule[-0.175pt]{4.818pt}{0.350pt}}
\put(777,561){\rule[-0.175pt]{4.818pt}{0.350pt}}
\put(962,586){\rule[-0.175pt]{0.350pt}{24.090pt}}
\put(952,586){\rule[-0.175pt]{4.818pt}{0.350pt}}
\put(952,686){\rule[-0.175pt]{4.818pt}{0.350pt}}
\sbox{\plotpoint}{\rule[-0.250pt]{0.500pt}{0.500pt}}%
\put(326,158){\usebox{\plotpoint}}
\put(335,176){\usebox{\plotpoint}}
\put(346,194){\usebox{\plotpoint}}
\put(358,210){\usebox{\plotpoint}}
\put(371,226){\usebox{\plotpoint}}
\put(385,242){\usebox{\plotpoint}}
\put(399,257){\usebox{\plotpoint}}
\put(414,271){\usebox{\plotpoint}}
\put(429,285){\usebox{\plotpoint}}
\put(445,299){\usebox{\plotpoint}}
\put(461,312){\usebox{\plotpoint}}
\put(477,325){\usebox{\plotpoint}}
\put(493,338){\usebox{\plotpoint}}
\put(510,351){\usebox{\plotpoint}}
\put(526,364){\usebox{\plotpoint}}
\put(543,376){\usebox{\plotpoint}}
\put(560,388){\usebox{\plotpoint}}
\put(576,401){\usebox{\plotpoint}}
\put(593,413){\usebox{\plotpoint}}
\put(610,425){\usebox{\plotpoint}}
\put(627,437){\usebox{\plotpoint}}
\put(644,449){\usebox{\plotpoint}}
\put(660,461){\usebox{\plotpoint}}
\put(677,473){\usebox{\plotpoint}}
\put(695,485){\usebox{\plotpoint}}
\put(711,497){\usebox{\plotpoint}}
\put(728,509){\usebox{\plotpoint}}
\put(745,521){\usebox{\plotpoint}}
\put(762,533){\usebox{\plotpoint}}
\put(779,545){\usebox{\plotpoint}}
\put(796,557){\usebox{\plotpoint}}
\put(813,568){\usebox{\plotpoint}}
\put(830,580){\usebox{\plotpoint}}
\put(847,592){\usebox{\plotpoint}}
\put(864,604){\usebox{\plotpoint}}
\put(882,615){\usebox{\plotpoint}}
\put(899,627){\usebox{\plotpoint}}
\put(916,639){\usebox{\plotpoint}}
\put(932,651){\usebox{\plotpoint}}
\put(950,663){\usebox{\plotpoint}}
\put(967,674){\usebox{\plotpoint}}
\put(984,686){\usebox{\plotpoint}}
\put(1001,698){\usebox{\plotpoint}}
\put(1018,710){\usebox{\plotpoint}}
\put(1035,721){\usebox{\plotpoint}}
\put(1053,733){\usebox{\plotpoint}}
\put(1070,745){\usebox{\plotpoint}}
\put(1087,756){\usebox{\plotpoint}}
\put(1104,768){\usebox{\plotpoint}}
\put(1121,780){\usebox{\plotpoint}}
\put(1131,787){\usebox{\plotpoint}}
\end{picture}

\setlength{\unitlength}{0.240900pt}
\ifx\plotpoint\undefined\newsavebox{\plotpoint}\fi
\sbox{\plotpoint}{\rule[-0.175pt]{0.350pt}{0.350pt}}%
\begin{picture}(1200,900)(0,0)
\tenrm
\sbox{\plotpoint}{\rule[-0.175pt]{0.350pt}{0.350pt}}%
\put(264,158){\rule[-0.175pt]{210.065pt}{0.350pt}}
\put(264,158){\rule[-0.175pt]{0.350pt}{151.526pt}}
\put(264,284){\rule[-0.175pt]{4.818pt}{0.350pt}}
\put(242,284){\makebox(0,0)[r]{$1$}}
\put(1116,284){\rule[-0.175pt]{4.818pt}{0.350pt}}
\put(264,410){\rule[-0.175pt]{4.818pt}{0.350pt}}
\put(242,410){\makebox(0,0)[r]{$2$}}
\put(1116,410){\rule[-0.175pt]{4.818pt}{0.350pt}}
\put(264,535){\rule[-0.175pt]{4.818pt}{0.350pt}}
\put(242,535){\makebox(0,0)[r]{$3$}}
\put(1116,535){\rule[-0.175pt]{4.818pt}{0.350pt}}
\put(264,661){\rule[-0.175pt]{4.818pt}{0.350pt}}
\put(242,661){\makebox(0,0)[r]{$4$}}
\put(1116,661){\rule[-0.175pt]{4.818pt}{0.350pt}}
\put(438,158){\rule[-0.175pt]{0.350pt}{4.818pt}}
\put(438,113){\makebox(0,0){$1$}}
\put(438,767){\rule[-0.175pt]{0.350pt}{4.818pt}}
\put(613,158){\rule[-0.175pt]{0.350pt}{4.818pt}}
\put(613,113){\makebox(0,0){$2$}}
\put(613,767){\rule[-0.175pt]{0.350pt}{4.818pt}}
\put(787,158){\rule[-0.175pt]{0.350pt}{4.818pt}}
\put(787,113){\makebox(0,0){$3$}}
\put(787,767){\rule[-0.175pt]{0.350pt}{4.818pt}}
\put(962,158){\rule[-0.175pt]{0.350pt}{4.818pt}}
\put(962,113){\makebox(0,0){$4$}}
\put(962,767){\rule[-0.175pt]{0.350pt}{4.818pt}}
\put(264,158){\rule[-0.175pt]{210.065pt}{0.350pt}}
\put(1136,158){\rule[-0.175pt]{0.350pt}{151.526pt}}
\put(264,787){\rule[-0.175pt]{210.065pt}{0.350pt}}
\put(45,472){\makebox(0,0)[l]{\shortstack{$a\,V(r)$}}}
\put(700,68){\makebox(0,0){$r/a$}}
\put(395,693){\makebox(0,0)[l]{b) Improved Action}}
\put(264,158){\rule[-0.175pt]{0.350pt}{151.526pt}}
\put(438,270){\circle{12}}
\put(511,332){\circle{12}}
\put(566,377){\circle{12}}
\put(613,402){\circle{12}}
\put(654,429){\circle{12}}
\put(757,496){\circle{12}}
\put(787,530){\circle{12}}
\put(962,669){\circle{12}}
\put(438,269){\usebox{\plotpoint}}
\put(428,269){\rule[-0.175pt]{4.818pt}{0.350pt}}
\put(428,270){\rule[-0.175pt]{4.818pt}{0.350pt}}
\put(511,332){\usebox{\plotpoint}}
\put(501,332){\rule[-0.175pt]{4.818pt}{0.350pt}}
\put(501,332){\rule[-0.175pt]{4.818pt}{0.350pt}}
\put(566,376){\rule[-0.175pt]{0.350pt}{0.482pt}}
\put(556,376){\rule[-0.175pt]{4.818pt}{0.350pt}}
\put(556,378){\rule[-0.175pt]{4.818pt}{0.350pt}}
\put(613,401){\rule[-0.175pt]{0.350pt}{0.482pt}}
\put(603,401){\rule[-0.175pt]{4.818pt}{0.350pt}}
\put(603,403){\rule[-0.175pt]{4.818pt}{0.350pt}}
\put(654,427){\rule[-0.175pt]{0.350pt}{0.964pt}}
\put(644,427){\rule[-0.175pt]{4.818pt}{0.350pt}}
\put(644,431){\rule[-0.175pt]{4.818pt}{0.350pt}}
\put(757,495){\rule[-0.175pt]{0.350pt}{0.482pt}}
\put(747,495){\rule[-0.175pt]{4.818pt}{0.350pt}}
\put(747,497){\rule[-0.175pt]{4.818pt}{0.350pt}}
\put(787,524){\rule[-0.175pt]{0.350pt}{3.132pt}}
\put(777,524){\rule[-0.175pt]{4.818pt}{0.350pt}}
\put(777,537){\rule[-0.175pt]{4.818pt}{0.350pt}}
\put(962,656){\rule[-0.175pt]{0.350pt}{6.022pt}}
\put(952,656){\rule[-0.175pt]{4.818pt}{0.350pt}}
\put(952,681){\rule[-0.175pt]{4.818pt}{0.350pt}}
\sbox{\plotpoint}{\rule[-0.250pt]{0.500pt}{0.500pt}}%
\put(339,158){\usebox{\plotpoint}}
\put(349,175){\usebox{\plotpoint}}
\put(362,192){\usebox{\plotpoint}}
\put(375,208){\usebox{\plotpoint}}
\put(388,223){\usebox{\plotpoint}}
\put(404,238){\usebox{\plotpoint}}
\put(419,252){\usebox{\plotpoint}}
\put(434,266){\usebox{\plotpoint}}
\put(450,279){\usebox{\plotpoint}}
\put(466,292){\usebox{\plotpoint}}
\put(482,305){\usebox{\plotpoint}}
\put(498,318){\usebox{\plotpoint}}
\put(515,331){\usebox{\plotpoint}}
\put(531,344){\usebox{\plotpoint}}
\put(548,356){\usebox{\plotpoint}}
\put(565,368){\usebox{\plotpoint}}
\put(582,380){\usebox{\plotpoint}}
\put(598,392){\usebox{\plotpoint}}
\put(615,405){\usebox{\plotpoint}}
\put(632,417){\usebox{\plotpoint}}
\put(649,429){\usebox{\plotpoint}}
\put(666,441){\usebox{\plotpoint}}
\put(683,453){\usebox{\plotpoint}}
\put(700,465){\usebox{\plotpoint}}
\put(717,477){\usebox{\plotpoint}}
\put(734,489){\usebox{\plotpoint}}
\put(751,501){\usebox{\plotpoint}}
\put(768,512){\usebox{\plotpoint}}
\put(785,524){\usebox{\plotpoint}}
\put(802,536){\usebox{\plotpoint}}
\put(819,548){\usebox{\plotpoint}}
\put(836,559){\usebox{\plotpoint}}
\put(853,571){\usebox{\plotpoint}}
\put(870,583){\usebox{\plotpoint}}
\put(887,595){\usebox{\plotpoint}}
\put(904,607){\usebox{\plotpoint}}
\put(921,618){\usebox{\plotpoint}}
\put(938,630){\usebox{\plotpoint}}
\put(956,642){\usebox{\plotpoint}}
\put(973,654){\usebox{\plotpoint}}
\put(990,665){\usebox{\plotpoint}}
\put(1007,677){\usebox{\plotpoint}}
\put(1024,689){\usebox{\plotpoint}}
\put(1041,700){\usebox{\plotpoint}}
\put(1059,712){\usebox{\plotpoint}}
\put(1075,724){\usebox{\plotpoint}}
\put(1093,735){\usebox{\plotpoint}}
\put(1110,747){\usebox{\plotpoint}}
\put(1127,759){\usebox{\plotpoint}}
\put(1136,765){\usebox{\plotpoint}}
\end{picture}

\end{center}
\caption{Static-quark potential computed on $6^4$ lattices with $a\approx
0.4$~fm using the $\beta=4.5$ Wilson action and the
improved action with $\beta_\pl = 6.8$.}
\label{potl}
\end{figure}
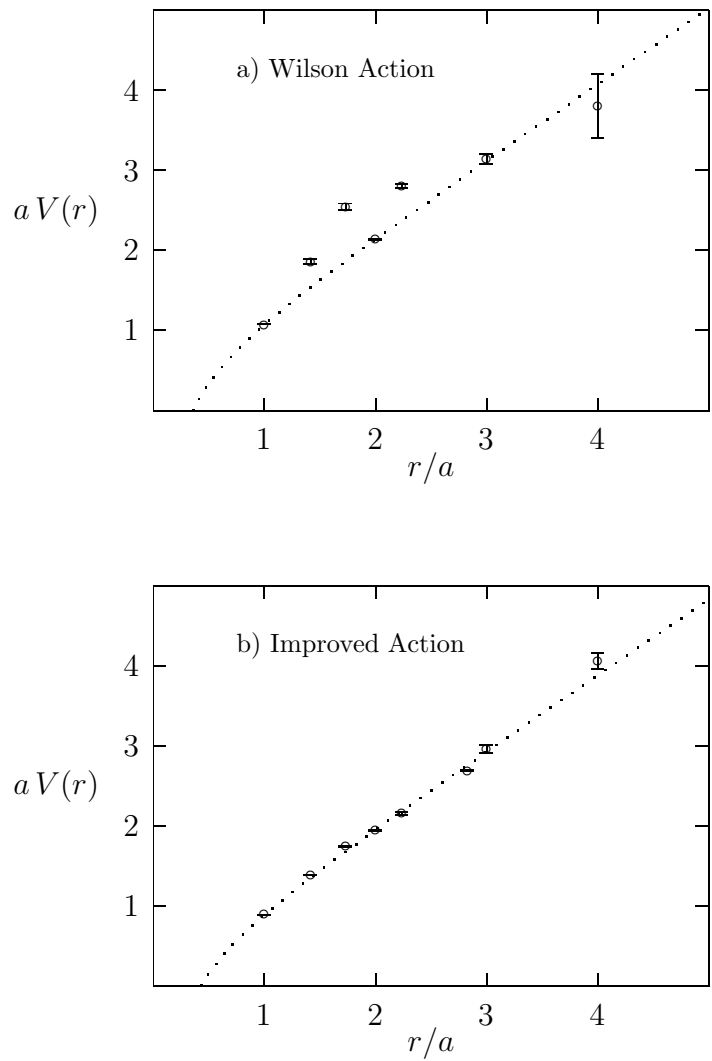

The standard Wilson action for gluons has finite-$a$ errors of order $a^2$.
On coarse lattices these lattice artifacts lead to severe (up to 40\%)
deviations from rotational invariance of
the static quark potential. This can be seen clearly in Fig.~\ref{potl}a
(the potential computed on a lattice with $a\approx 0.4$~fm), where the
points for $r/a =\sqrt{2}$ and $\sqrt{3}$ lie far off the line defined by
$r/a=1,2,3$.
These artifacts arise because
the plaquette operator, from which the
Wilson action is constructed, contains
$\order(a^2)$ terms
beyond the desired gluon kinetic term when it is expanded in
powers of derivatives of the gauge field\,\cite{Wei83}:
\beq\label{expansion}
\ba{r@{~}c@{~}l}
1-\third \Re\Tr U_\pl &=&
  \displaystyle r_0^{(\pl)} \sum_{\mu,\nu} \Tr(F_{\mu\nu} F_{\mu\nu})
  + a^2\Bigl[ r_1^{(\pl)} R_1 + r_2^{(\pl)} R_2 + r_3^{(\pl)} R_3 \Bigr]
  \\[2.5ex]
&+& \order(a^4) + \hbox{total derivatives},
\ea
\eeq
where
\beq
\ba{r@{~}c@{~}l}
R_1 &=& \sum_{\mu,\nu}\Tr(D_\mu F_{\mu\nu} D_\mu F_{\mu\nu}), \\[2.5ex]
R_2 &=& \sum_{\mu,\nu,\si} \Tr(D_\mu F_{\nu\si} D_\mu F_{\nu\si}), \\[2.5ex]
R_3 &=& \sum_{\mu,\nu,\si} \Tr(D_\mu F_{\mu\si} D_\nu F_{\nu\si}).
\ea
\eeq
Here the $r_i$ are coefficients in the operator product expansion of
the plaquette. (Tree-level $r_i$'s are tabulated for a variety of loop
operators, like the plaquette, in~\cite{lw}.)
Note that $R_1$ communicates the lattice's violation of Lorentz invariance,
and is therefore responsible for the bad behavior of the static potential,
while $R_2$ and $R_3$ are Lorentz invariant.
If we want to eliminate the $\order(a^2)$ lattice artifacts then we
need to form an improved action by adding other Wilson loops to the action,
which will give canceling amounts of $R_1, R_2, R_3$.

Only $R_1$ contributes in \Eqn{expansion} at tree-level,
but quantum corrections bring in the
other two operators. To remove all three,
it might seem that we need to add three new Wilson loops to cancel these
terms, but actually the coefficient of $R_3$
can be set to zero by a change of field
variable in the path integral,
\beq \label{Atransform}
A_\mu \to A_\mu + a^2\,\alpha_s\,f(\alpha_s)\,\sum_\nu D_\nu F_{\nu\mu},
\eeq
so only two new terms are needed\,\cite{Cur83,lw}.
There are many possible choices
for these, but we take the rectangle and ``parallelogram'':
\beq
U_\rt =
\begin{picture}(60,30)(0,15)
  \put(10,10){\vector(0,1){12.5}}
  \put(10,10){\line(0,1){20}}
  \put(10,30){\vector(1,0){12.5}}
  \put(10,30){\vector(1,0){32.5}}
  \put(10,30){\line(1,0){40}}
  \put(50,30){\vector(0,-1){12.5}}
  \put(50,30){\line(0,-1){20}}
  \put(50,10){\vector(-1,0){12.5}}
  \put(50,10){\vector(-1,0){32.5}}
  \put(50,10){\line(-1,0){40}}
\end{picture},
\quad
U_\pg =
\begin{picture}(60,30)(0,15)
  \put(10,10){\vector(0,1){12.5}}
  \put(10,10){\line(0,1){20}}
  \put(10,30){\vector(2,1){10}}
  \put(10,30){\line(2,1){15}}
  \put(25.2,37.6){\vector(1,0){12.5}}
  \put(25.2,37.6){\line(1,0){20}}
  \put(45.2,37.6){\vector(0,-1){12.5}}
  \put(45.2,37.6){\line(0,-1){20}}
  \put(45.2,17.6){\vector(-2,-1){10}}
  \put(45.2,17.6){\line(-2,-1){15}}
  \put(30,10){\vector(-1,0){12.5}}
  \put(30,10){\line(-1,0){20}}
\end{picture}.
\eeq
The improved action is\,\cite{Cur83,lw}
\bearray
S[U] &=& \beta_\pl \sum_\pl \ls{U_\pl} \nl
&+& \beta_\rt \sum_\rt \ls{U_\rt} \nl
&+& \beta_\pg \sum_\pg \ls{U_\pg} ,
\eearray
with $\beta_\pl$ given as an input, and $\be_\rt$ and $\be_\pg$
computed in tadpole-improved perturbation theory to cancel out
the $\order(a^2)$ terms in the derivative expansion of the action.
At tree-level, the~$\beta$'s are readily computed by combining expansions like
\Eqn{expansion} for each of the three loops.
They are tadpole-improved by dividing each
Wilson loop with $L$ links by $(u_0)^L$\,\cite{lm}. One-loop
corrections have also been computed\,\cite{w}, but must be adjusted to
account for the tadpole improvement. We find:
\bearray
\label{oneloopi}
\beta_\rt &=& -\frac{\beta_\pl}{20\,u_0^2}\, \left( 1 + 0.4805\,\alpha_s
    \right), \\
\label{oneloopii}
\beta_\pg &=& -\frac{\beta_\pl}{u_0^2} \, 0.03325\,\alpha_s.
\eearray

Following~\cite{lm}, we use the measured
expectation value of the plaquette to determine both the value of the mean
link~$u_0$ and the QCD coupling constant~$\alpha_s$,
\bearray
 u_0 &=& \left(\mbox{$\frac{1}{3}$\,Re\,Tr}
   \langle U_\pl \rangle\right)^{1/4}, \\
 \alpha_s &=& -\frac{\ln\Bigl({\textstyle {1\over 3}}\mbox{Re\,Tr\,}
   \langle U_\pl\rangle\Bigr)}{3.06839}.
\eearray
This result follows from the tree-level perturbative
calculation of the plaquette\,\cite{ww}.
The couplings $\beta_\rt$ and $\beta_\pg$ are determined self-consistently with
$u_0$ and~$\alpha_s$ for a given~$\beta_\pl$. As in NRQCD, there is no tuning
of the couplings for the correction terms: tadpole-improved perturbation
theory determines them in terms of the single bare
coupling~$\beta_\pl$. Using identities from~\cite{lw} we find that our
action is positive semidefinite at least for $\beta_\pl\ge6.8$,
which is necessary if perturbation theory is to be reliable.

The perturbative expansion \eqn{oneloopi} for $\be_\rt$ already demonstrates
the power of tadpole improvement. If one omits the tadpole factor
$u_0^2$, the expression becomes $\be_\rt = -\be_\pl(1+2.0\,\al_s)/20$.
Note that the coefficient of the one-loop-term $\al_s$ has quadrupled.
Tadpole improvement automatically supplies 75\% of the one-loop
contribution needed without improvement.
Since $\al_s\approx 0.3$, the unimproved expansion is not particularly
convergent. However, with tadpole improvement, the one-loop correction
is only about 10--20\% of~$\beta_\rt$.

As indicated above, our improved action is not
unique. Our techniques should work as well for actions with other
forms for the correction terms. To verify this, we compare results obtained
using our improved action above with those obtained from a
tree-level improved action with a very different correction
term\,\cite{Curtrt}:
\bearray \label{trtS}
S_\trt[U] &=& \beta_\pl \sum_\pl \ls{U_\pl} \nl
&+& \beta_\trt \sum_\trt \ls{U_\trt},
\eearray
where $U_\trt$ is a twisted rectangle operator,
\beq
U_{\trt} =
\begin{picture}(60,30)(0,15)

  \put(27.5,10){\vector(-1,0){10}}
  \put(27.5,10){\line(-1,0){17.5}}
  \put(10,10){\vector(0,1){12.5}}
  \put(10,10){\line(0,1){20}}
  \put(10,30){\vector(1,0){12.5}}
  \put(10,30){\line(1,0){17.5}}

  \put(32.5,10){\vector(1,0){10}}
  \put(32.5,10){\line(1,0){17.5}}
  \put(50,10){\vector(0,1){12.5}}
  \put(50,10){\line(0,1){20}}
  \put(50,30){\vector(-1,0){12.5}}
  \put(50,30){\line(-1,0){17.5}}

  \put(32.5,10){\line(-1,4){5.0}}
  \put(27.5,10){\line( 1,4){5.0}}

\end{picture}.
\eeq
At tree-level, with tadpole improvement,
\beq
\beta_\trt = \frac{\beta_\pl}{12\, u_0^4}.
\eeq
Note that tadpole improvement introduces four powers of $u_0$ here,
rather than the two powers in $\beta_\rt$ above. This makes $S_\trt$
much more sensitive to tadpole improvement.

\section{Monte-Carlo Results}

We conducted simulations with improved actions for a range of large
lattice spacings (Table~\ref{params}).
The static-quark potential computed using our improved gluon action
and the coarsest lattice is
shown in Fig.~\ref{potl}b. As in the Wilson case (Fig~\ref{potl}a),
the lattice spacing is about 0.4~fm.  The dashed line in these plots
is the standard infrared parameterization for the continuum potential,
$V(r)=Kr-\pi/12r + c$, adjusted to fit the on-axis values of the
potential.
Off-axis points deviate from the fit by  40\% for
the Wilson theory, indicating a significant failure of rotation
invariance due to finite-$a$ errors. By contrast, the deviations are
only~2--4\% for the improved theory\,---\,negligible for most
low-energy applications.

\begin{table}
\begin{center}
\begin{tabular}{lccccc} \hline \\
Action & $\beta_\pl$ & $\langle\frac{1}{3}$\,Re\,Tr$U_\pl\rangle$
& $aM_c^0$ & $a$ & dimensions \\
\hline
improved & 6.8 & .46 & 2.40 & .40 fm & $6^4$, $6^3\times 9$ \\
& 7.1 & .506 & 1.92 & .33 fm & $7^3\times 10$ \\
& 7.4 & .56 & 1.30 & .24 fm & $8^3\times 10$, $9^3\times 12$ \\
\\
$S_\trt$ & 4.1 & .454 & 2.40 & .40 fm & $6^3\times 9$ \\ \\
Wilson & 4.5 & .34 & 2.00 & .40 fm & $6^4$ \\
& 5.7 & .55 & \dgsp .80 & .17 fm & $12^3\times 24$, $16^4$ \\
\hline \hline
\end{tabular}
\end{center}
\caption{Parameters used in our simulations.
The additional couplings for improved actions were determined using
the values listed in the table for  $\beta_\pl$ and the plaquette.
$M_c^0$ is the bare charm-quark mass used in the NRQCD quark action.}
\label{params}
\end{table}

To assess the relative importance of tree-level improvement, tadpole
improvement and one-loop corrections we computed the potential for
several different actions, all with lattice spacings of about
0.4~fm. We focused on the deviation $\Delta V$ of $V(a,a,a)$ from the continuum
potential adjusted to fit on-axis values of the simulated potential.
$\Delta V$~is a sensitive indicator of violations of rotational invariance.
Our results are in Table~\ref{dV}. As expected, the correction
term in the action is significantly underestimated without tadpole
improvement. The tadpole-improved action is very accurate both with
and without one-loop corrections, suggesting that $\order(a^2\alpha_s)$
corrections are comparable to those of $\order(a^4)$.

We have also included in Table~\ref{dV} results obtained using the
twisted-rectangle action $S_\trt$~\eqn{trtS}. We find that this action gives
a potential that is essentially identical to that obtained with our other
improved action. Note that tadpole improvement more than doubles the
size of the correction term in $S_\trt$ when $a= 0.4$~fm. The
quality of the results obtained from $S_\trt$ is strong evidence in
support of tadpole improvement.

\begin{table}
\begin{center}
\begin{tabular}{lc} \hline
\\
Action &  $\Delta V(\sqrt{3} a)/ K\sqrt{3} a$ \nl \hline
unimproved (Wilson) & .41 (2) \nl
\mbox{} \nl
tree-level improved, no tadpole improvement & .15 (1) \nl
one-loop improved, no tadpole improvement & .12 (2) \nl
\mbox{} \nl
tree-level improved, with tadpole improvement & .05 (1) \nl
one-loop improved, with tadpole improvement & .04 (1) \nl
\mbox{} \nl
twisted-rectangle correction, with tadpole improvement &
 .04 (2)\nl \hline\hline
\end{tabular}
\end{center}
\caption{Error in the static quark potential at $V(a,a,a)$ for a variety
of gluon actions. The lattice spacing in each case is $a\approx
0.4$~fm; $K$ is the slope of the linear part of the static potential.}
\label{dV}
\end{table}

In computing the potentials, we assumed that the static quark propagator
is simply the product of link operators $U_\mu(x)$ along the time axis.
Our action is designed so that this is true at tree level through
$\order(a^2)$, however there are corrections of order $\al_s a^2$.
These arise because of the field transformation in \Eqn{Atransform}.
Our potential therefore has errors of $\order(\alpha_s a^2)$, even
though the action is accurate up to errors of $\order(\al_s^2 a^2,a^4)$. Note
that these particular errors in our potential
do not break rotational invariance and so
have no effect on the values of $\Delta V$ in Table~\ref{dV}.
A straightforward perturbative calculation
is needed to remove the $\order(\al_s a^2)$ errors from our potential.

To further check on our improved theory, we examined the spin-averaged
spectrum of the $\psi$ family of mesons using NRQCD for the $c$-quarks
and our improved action at $\beta_\pl=$~6.8, 7.1 and~7.4. Because we were
examining only the spin-averaged spectrum, we omitted all spin-dependent
corrections from the NRQCD action, but kept the corrections for
$\order(a,a^2)$~errors, and for spin-independent $\order(v^2/c^2)$ effects
\,\cite{nrqcd}.
The spectra, normalized to give the correct
$1P$-$1S$~splitting, are shown in Fig.~\ref{spect} together with
experimental results (dashed lines) and simulation results obtained
using the Wilson action for the gluons at
smaller~$a$'s\,\cite{nrqcd}. All agree to within 30--40~MeV. (The
$2S$~simulation masses are a little above the true mass, which is
expected since the simulation does not include light-quark loops.)
\begin{figure}
\begin{center}
\setlength{\unitlength}{.02in}
\begin{picture}(120,130)(0,280)
\put(89,340){\circle{3}}\put(95,340){\makebox(0,0)[l]{$0.40$~fm}}
\put(88,329){\framebox(2,2){\mbox{}}}
             \put(95,330){\makebox(0,0)[l]{$0.33$~fm}}
\put(89,320){{\circle*{3}}}\put(95,320){\makebox(0,0)[l]{$0.24$~fm}}
\put(88,310){\rule[-\unitlength]{2\unitlength}{2\unitlength}}
         \put(95,310){\makebox(0,0)[l]{$0.17$~fm}}

\put(15,290){\line(0,1){120}}
\multiput(13,300)(0,50){3}{\line(1,0){4}}
\multiput(14,310)(0,10){9}{\line(1,0){2}}
\put(12,300){\makebox(0,0)[r]{3.0}}
\put(12,350){\makebox(0,0)[r]{3.5}}
\put(12,400){\makebox(0,0)[r]{4.0}}
\put(12,410){\makebox(0,0)[r]{GeV}}

\put(30,290){\makebox(0,0)[t]{$S$}}

\multiput(23,307)(3,0){6}{\line(1,0){2}}
\put(26,307){\circle{3}}
\put(29,306){\framebox(2,2){\mbox{}}}
\put(34,307){\circle*{3}}
\put(37,306){\rule{2\unitlength}{2\unitlength}}

\multiput(23,366)(3,0){6}{\line(1,0){2}}
\put(26,371){\circle{3}}
\put(26,369){\line(0,1){4}}
\put(29,371){\framebox(2,2){\mbox{}}}
\put(30,370){\line(0,1){4}}
\put(34,372){\circle*{3}}
\put(34,368){\line(0,1){8}}
\put(37,369){\rule{2\unitlength}{2\unitlength}}
\put(38,361){\line(0,1){16}}

\put(50,290){\makebox(0,0)[t]{$P$}}

\multiput(43,352)(3,0){6}{\line(1,0){2}}
\put(46,352){\circle{3}}
\put(49,351){\framebox(2,2){}}
\put(54,352){\circle*{3}}
\put(54,351){\line(0,1){2}}
\put(57,351){\rule{2\unitlength}{2\unitlength}}
\put(58,351.5){\line(0,1){1}}

\put(70,290){\makebox(0,0)[t]{$D$}}

\put(66,387){\circle{3}}
\put(66,382){\line(0,1){10}}
\put(69,382){\framebox(2,2){}}
\put(70,376){\line(0,1){12}}
\put(74,387){\circle*{3}}
\put(74,384){\line(0,1){6}}
\put(77,382){\rule{2\unitlength}{2\unitlength}}
\put(78,378){\line(0,1){10}}
\end{picture}


\end{center}
\caption{$S$, $P$, and $D$ states of charmonium computed on lattices with:
$a=0.40$~fm (improved action, $\beta_\pl=6.8$);
$a=0.33$~fm (improved action, $\beta_\pl=7.1$);
$a=0.24$~fm (improved action, $\beta_\pl=7.4$); and
$a=0.17$~fm (Wilson action, $\beta=5.7$, from~[6]). The dashed lines
indicate the true masses.}
\label{spect}
\end{figure}
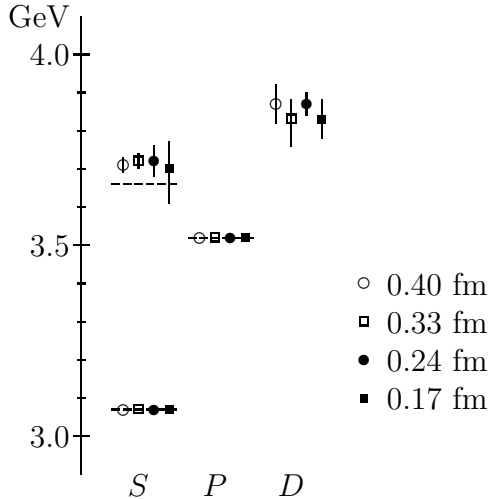

\begin{figure}
\begin{centering}
\setlength{\unitlength}{0.240900pt}
\ifx\plotpoint\undefined\newsavebox{\plotpoint}\fi
\sbox{\plotpoint}{\rule[-0.175pt]{0.350pt}{0.350pt}}%
\begin{picture}(1200,900)(0,0)
\tenrm
\sbox{\plotpoint}{\rule[-0.175pt]{0.350pt}{0.350pt}}%
\put(264,158){\rule[-0.175pt]{210.065pt}{0.350pt}}
\put(264,158){\rule[-0.175pt]{0.350pt}{151.526pt}}
\put(264,420){\rule[-0.175pt]{4.818pt}{0.350pt}}
\put(242,420){\makebox(0,0)[r]{$5$}}
\put(1116,420){\rule[-0.175pt]{4.818pt}{0.350pt}}
\put(264,682){\rule[-0.175pt]{4.818pt}{0.350pt}}
\put(242,682){\makebox(0,0)[r]{$10$}}
\put(1116,682){\rule[-0.175pt]{4.818pt}{0.350pt}}
\put(264,158){\rule[-0.175pt]{0.350pt}{4.818pt}}
\put(264,113){\makebox(0,0){$0$}}
\put(264,767){\rule[-0.175pt]{0.350pt}{4.818pt}}
\put(700,158){\rule[-0.175pt]{0.350pt}{4.818pt}}
\put(700,113){\makebox(0,0){$0.5$}}
\put(700,767){\rule[-0.175pt]{0.350pt}{4.818pt}}
\put(1136,158){\rule[-0.175pt]{0.350pt}{4.818pt}}
\put(1136,113){\makebox(0,0){$1$}}
\put(1136,767){\rule[-0.175pt]{0.350pt}{4.818pt}}
\put(264,158){\rule[-0.175pt]{210.065pt}{0.350pt}}
\put(1136,158){\rule[-0.175pt]{0.350pt}{151.526pt}}
\put(264,787){\rule[-0.175pt]{210.065pt}{0.350pt}}
\put(45,472){\makebox(0,0)[l]{\shortstack{$R_{1S}(r)$}}}
\put(700,68){\makebox(0,0){$r$ (fm)}}
\put(700,832){\makebox(0,0){$1S$ Radial Wavefunction}}
\put(264,158){\rule[-0.175pt]{0.350pt}{151.526pt}}
\put(1006,722){\makebox(0,0)[r]{$a=0.40$~fm}}
\put(1050,722){\circle*{18}}
\put(867,200){\circle*{18}}
\put(1116,167){\circle*{18}}
\put(264,619){\circle*{18}}
\put(612,348){\circle*{18}}
\put(960,189){\circle*{18}}
\put(756,243){\circle*{18}}
\put(1042,175){\circle*{18}}
\put(1006,677){\makebox(0,0)[r]{$a=0.24$~fm}}
\put(1050,677){\circle{18}}
\put(861,199){\circle{18}}
\put(1025,172){\circle{18}}
\put(630,319){\circle{18}}
\put(781,226){\circle{18}}
\put(964,179){\circle{18}}
\put(897,190){\circle{18}}
\put(1054,169){\circle{18}}
\put(264,731){\circle{18}}
\put(475,512){\circle{18}}
\put(686,284){\circle{18}}
\put(897,192){\circle{18}}
\put(1108,172){\circle{18}}
\put(562,390){\circle{18}}
\put(736,249){\circle{18}}
\put(931,184){\circle{18}}
\put(1134,169){\circle{18}}
\put(995,174){\circle{18}}
\put(1134,164){\circle{18}}
\sbox{\plotpoint}{\rule[-0.250pt]{0.500pt}{0.500pt}}%
\put(1006,632){\makebox(0,0)[r]{quark model}}
\put(1028,632){\usebox{\plotpoint}}
\put(1048,632){\usebox{\plotpoint}}
\put(1069,632){\usebox{\plotpoint}}
\put(1090,632){\usebox{\plotpoint}}
\put(1094,632){\usebox{\plotpoint}}
\put(264,738){\usebox{\plotpoint}}
\put(264,738){\usebox{\plotpoint}}
\put(279,724){\usebox{\plotpoint}}
\put(295,711){\usebox{\plotpoint}}
\put(309,696){\usebox{\plotpoint}}
\put(324,680){\usebox{\plotpoint}}
\put(338,665){\usebox{\plotpoint}}
\put(352,650){\usebox{\plotpoint}}
\put(365,634){\usebox{\plotpoint}}
\put(378,617){\usebox{\plotpoint}}
\put(391,601){\usebox{\plotpoint}}
\put(404,585){\usebox{\plotpoint}}
\put(417,569){\usebox{\plotpoint}}
\put(429,553){\usebox{\plotpoint}}
\put(442,536){\usebox{\plotpoint}}
\put(455,520){\usebox{\plotpoint}}
\put(468,504){\usebox{\plotpoint}}
\put(481,487){\usebox{\plotpoint}}
\put(494,471){\usebox{\plotpoint}}
\put(507,455){\usebox{\plotpoint}}
\put(521,440){\usebox{\plotpoint}}
\put(534,424){\usebox{\plotpoint}}
\put(549,409){\usebox{\plotpoint}}
\put(563,394){\usebox{\plotpoint}}
\put(577,379){\usebox{\plotpoint}}
\put(592,364){\usebox{\plotpoint}}
\put(607,350){\usebox{\plotpoint}}
\put(622,336){\usebox{\plotpoint}}
\put(638,323){\usebox{\plotpoint}}
\put(654,309){\usebox{\plotpoint}}
\put(671,297){\usebox{\plotpoint}}
\put(688,285){\usebox{\plotpoint}}
\put(705,273){\usebox{\plotpoint}}
\put(722,262){\usebox{\plotpoint}}
\put(740,252){\usebox{\plotpoint}}
\put(759,242){\usebox{\plotpoint}}
\put(777,233){\usebox{\plotpoint}}
\put(797,225){\usebox{\plotpoint}}
\put(816,217){\usebox{\plotpoint}}
\put(835,210){\usebox{\plotpoint}}
\put(855,204){\usebox{\plotpoint}}
\put(875,198){\usebox{\plotpoint}}
\put(895,193){\usebox{\plotpoint}}
\put(915,188){\usebox{\plotpoint}}
\put(936,184){\usebox{\plotpoint}}
\put(956,180){\usebox{\plotpoint}}
\put(977,177){\usebox{\plotpoint}}
\put(997,174){\usebox{\plotpoint}}
\put(1018,172){\usebox{\plotpoint}}
\put(1038,170){\usebox{\plotpoint}}
\put(1059,168){\usebox{\plotpoint}}
\put(1080,166){\usebox{\plotpoint}}
\put(1101,165){\usebox{\plotpoint}}
\put(1121,164){\usebox{\plotpoint}}
\put(1136,164){\usebox{\plotpoint}}
\end{picture}

\setlength{\unitlength}{0.240900pt}
\ifx\plotpoint\undefined\newsavebox{\plotpoint}\fi
\sbox{\plotpoint}{\rule[-0.175pt]{0.350pt}{0.350pt}}%
\begin{picture}(1200,900)(0,0)
\tenrm
\sbox{\plotpoint}{\rule[-0.175pt]{0.350pt}{0.350pt}}%
\put(264,158){\rule[-0.175pt]{210.065pt}{0.350pt}}
\put(264,158){\rule[-0.175pt]{0.350pt}{151.526pt}}
\put(264,368){\rule[-0.175pt]{4.818pt}{0.350pt}}
\put(242,368){\makebox(0,0)[r]{$2$}}
\put(1116,368){\rule[-0.175pt]{4.818pt}{0.350pt}}
\put(264,577){\rule[-0.175pt]{4.818pt}{0.350pt}}
\put(242,577){\makebox(0,0)[r]{$4$}}
\put(1116,577){\rule[-0.175pt]{4.818pt}{0.350pt}}
\put(264,787){\rule[-0.175pt]{4.818pt}{0.350pt}}
\put(242,787){\makebox(0,0)[r]{$6$}}
\put(1116,787){\rule[-0.175pt]{4.818pt}{0.350pt}}
\put(264,158){\rule[-0.175pt]{0.350pt}{4.818pt}}
\put(264,113){\makebox(0,0){$0$}}
\put(264,767){\rule[-0.175pt]{0.350pt}{4.818pt}}
\put(700,158){\rule[-0.175pt]{0.350pt}{4.818pt}}
\put(700,113){\makebox(0,0){$0.5$}}
\put(700,767){\rule[-0.175pt]{0.350pt}{4.818pt}}
\put(1136,158){\rule[-0.175pt]{0.350pt}{4.818pt}}
\put(1136,113){\makebox(0,0){$1$}}
\put(1136,767){\rule[-0.175pt]{0.350pt}{4.818pt}}
\put(264,158){\rule[-0.175pt]{210.065pt}{0.350pt}}
\put(1136,158){\rule[-0.175pt]{0.350pt}{151.526pt}}
\put(264,787){\rule[-0.175pt]{210.065pt}{0.350pt}}
\put(45,472){\makebox(0,0)[l]{\shortstack{$R_{1P}(r)$}}}
\put(700,68){\makebox(0,0){$r$ (fm)}}
\put(700,832){\makebox(0,0){$1P$ Radial Wavefunction}}
\put(264,158){\rule[-0.175pt]{0.350pt}{151.526pt}}
\put(1006,722){\makebox(0,0)[r]{$a=0.40$~fm}}
\put(1050,722){\circle*{18}}
\put(612,521){\circle*{18}}
\put(756,411){\circle*{18}}
\put(1042,243){\circle*{18}}
\put(867,321){\circle*{18}}
\put(1116,214){\circle*{18}}
\put(960,299){\circle*{18}}
\put(1042,244){\circle*{18}}
\put(1116,211){\circle*{18}}
\put(1028,722){\rule[-0.175pt]{15.899pt}{0.350pt}}
\put(1028,712){\rule[-0.175pt]{0.350pt}{4.818pt}}
\put(1094,712){\rule[-0.175pt]{0.350pt}{4.818pt}}
\put(612,503){\rule[-0.175pt]{0.350pt}{8.431pt}}
\put(602,503){\rule[-0.175pt]{4.818pt}{0.350pt}}
\put(602,538){\rule[-0.175pt]{4.818pt}{0.350pt}}
\put(756,404){\rule[-0.175pt]{0.350pt}{3.613pt}}
\put(746,404){\rule[-0.175pt]{4.818pt}{0.350pt}}
\put(746,419){\rule[-0.175pt]{4.818pt}{0.350pt}}
\put(1042,238){\rule[-0.175pt]{0.350pt}{2.650pt}}
\put(1032,238){\rule[-0.175pt]{4.818pt}{0.350pt}}
\put(1032,249){\rule[-0.175pt]{4.818pt}{0.350pt}}
\put(867,314){\rule[-0.175pt]{0.350pt}{3.373pt}}
\put(857,314){\rule[-0.175pt]{4.818pt}{0.350pt}}
\put(857,328){\rule[-0.175pt]{4.818pt}{0.350pt}}
\put(1116,211){\rule[-0.175pt]{0.350pt}{1.445pt}}
\put(1106,211){\rule[-0.175pt]{4.818pt}{0.350pt}}
\put(1106,217){\rule[-0.175pt]{4.818pt}{0.350pt}}
\put(960,289){\rule[-0.175pt]{0.350pt}{4.577pt}}
\put(950,289){\rule[-0.175pt]{4.818pt}{0.350pt}}
\put(950,308){\rule[-0.175pt]{4.818pt}{0.350pt}}
\put(1042,240){\rule[-0.175pt]{0.350pt}{1.927pt}}
\put(1032,240){\rule[-0.175pt]{4.818pt}{0.350pt}}
\put(1032,248){\rule[-0.175pt]{4.818pt}{0.350pt}}
\put(1116,208){\rule[-0.175pt]{0.350pt}{1.686pt}}
\put(1106,208){\rule[-0.175pt]{4.818pt}{0.350pt}}
\put(1106,215){\rule[-0.175pt]{4.818pt}{0.350pt}}
\put(1006,677){\makebox(0,0)[r]{$a=0.24$~fm}}
\put(1050,677){\circle{18}}
\put(1054,231){\circle{18}}
\put(897,286){\circle{18}}
\put(931,265){\circle{18}}
\put(1025,222){\circle{18}}
\put(995,251){\circle{18}}
\put(1134,205){\circle{18}}
\put(475,555){\circle{18}}
\put(562,560){\circle{18}}
\put(736,442){\circle{18}}
\put(931,295){\circle{18}}
\put(964,248){\circle{18}}
\put(1134,246){\circle{18}}
\put(1054,214){\circle{18}}
\put(630,521){\circle{18}}
\put(781,399){\circle{18}}
\put(1108,164){\circle{18}}
\put(964,275){\circle{18}}
\put(1134,162){\circle{18}}
\put(1134,194){\circle{18}}
\put(686,480){\circle{18}}
\put(736,430){\circle{18}}
\put(861,329){\circle{18}}
\put(897,312){\circle{18}}
\put(1025,242){\circle{18}}
\put(1054,238){\circle{18}}
\put(781,389){\circle{18}}
\put(897,304){\circle{18}}
\put(1028,677){\rule[-0.175pt]{15.899pt}{0.350pt}}
\put(1028,667){\rule[-0.175pt]{0.350pt}{4.818pt}}
\put(1094,667){\rule[-0.175pt]{0.350pt}{4.818pt}}
\put(1054,221){\rule[-0.175pt]{0.350pt}{4.577pt}}
\put(1044,221){\rule[-0.175pt]{4.818pt}{0.350pt}}
\put(1044,240){\rule[-0.175pt]{4.818pt}{0.350pt}}
\put(897,264){\rule[-0.175pt]{0.350pt}{10.840pt}}
\put(887,264){\rule[-0.175pt]{4.818pt}{0.350pt}}
\put(887,309){\rule[-0.175pt]{4.818pt}{0.350pt}}
\put(931,254){\rule[-0.175pt]{0.350pt}{5.300pt}}
\put(921,254){\rule[-0.175pt]{4.818pt}{0.350pt}}
\put(921,276){\rule[-0.175pt]{4.818pt}{0.350pt}}
\put(1025,212){\rule[-0.175pt]{0.350pt}{4.818pt}}
\put(1015,212){\rule[-0.175pt]{4.818pt}{0.350pt}}
\put(1015,232){\rule[-0.175pt]{4.818pt}{0.350pt}}
\put(995,236){\rule[-0.175pt]{0.350pt}{6.986pt}}
\put(985,236){\rule[-0.175pt]{4.818pt}{0.350pt}}
\put(985,265){\rule[-0.175pt]{4.818pt}{0.350pt}}
\put(1134,197){\rule[-0.175pt]{0.350pt}{3.854pt}}
\put(1124,197){\rule[-0.175pt]{4.818pt}{0.350pt}}
\put(1124,213){\rule[-0.175pt]{4.818pt}{0.350pt}}
\put(475,483){\rule[-0.175pt]{0.350pt}{34.690pt}}
\put(465,483){\rule[-0.175pt]{4.818pt}{0.350pt}}
\put(465,627){\rule[-0.175pt]{4.818pt}{0.350pt}}
\put(562,522){\rule[-0.175pt]{0.350pt}{18.308pt}}
\put(552,522){\rule[-0.175pt]{4.818pt}{0.350pt}}
\put(552,598){\rule[-0.175pt]{4.818pt}{0.350pt}}
\put(736,407){\rule[-0.175pt]{0.350pt}{17.104pt}}
\put(726,407){\rule[-0.175pt]{4.818pt}{0.350pt}}
\put(726,478){\rule[-0.175pt]{4.818pt}{0.350pt}}
\put(931,265){\rule[-0.175pt]{0.350pt}{14.213pt}}
\put(921,265){\rule[-0.175pt]{4.818pt}{0.350pt}}
\put(921,324){\rule[-0.175pt]{4.818pt}{0.350pt}}
\put(964,237){\rule[-0.175pt]{0.350pt}{5.300pt}}
\put(954,237){\rule[-0.175pt]{4.818pt}{0.350pt}}
\put(954,259){\rule[-0.175pt]{4.818pt}{0.350pt}}
\put(1134,207){\rule[-0.175pt]{0.350pt}{18.790pt}}
\put(1124,207){\rule[-0.175pt]{4.818pt}{0.350pt}}
\put(1124,285){\rule[-0.175pt]{4.818pt}{0.350pt}}
\put(1054,207){\rule[-0.175pt]{0.350pt}{3.373pt}}
\put(1044,207){\rule[-0.175pt]{4.818pt}{0.350pt}}
\put(1044,221){\rule[-0.175pt]{4.818pt}{0.350pt}}
\put(630,483){\rule[-0.175pt]{0.350pt}{18.067pt}}
\put(620,483){\rule[-0.175pt]{4.818pt}{0.350pt}}
\put(620,558){\rule[-0.175pt]{4.818pt}{0.350pt}}
\put(781,375){\rule[-0.175pt]{0.350pt}{11.563pt}}
\put(771,375){\rule[-0.175pt]{4.818pt}{0.350pt}}
\put(771,423){\rule[-0.175pt]{4.818pt}{0.350pt}}
\put(1108,158){\rule[-0.175pt]{0.350pt}{6.022pt}}
\put(1098,158){\rule[-0.175pt]{4.818pt}{0.350pt}}
\put(1098,183){\rule[-0.175pt]{4.818pt}{0.350pt}}
\put(964,255){\rule[-0.175pt]{0.350pt}{9.636pt}}
\put(954,255){\rule[-0.175pt]{4.818pt}{0.350pt}}
\put(954,295){\rule[-0.175pt]{4.818pt}{0.350pt}}
\put(1134,158){\rule[-0.175pt]{0.350pt}{3.613pt}}
\put(1124,158){\rule[-0.175pt]{4.818pt}{0.350pt}}
\put(1124,173){\rule[-0.175pt]{4.818pt}{0.350pt}}
\put(1134,185){\rule[-0.175pt]{0.350pt}{4.095pt}}
\put(1124,185){\rule[-0.175pt]{4.818pt}{0.350pt}}
\put(1124,202){\rule[-0.175pt]{4.818pt}{0.350pt}}
\put(686,440){\rule[-0.175pt]{0.350pt}{19.031pt}}
\put(676,440){\rule[-0.175pt]{4.818pt}{0.350pt}}
\put(676,519){\rule[-0.175pt]{4.818pt}{0.350pt}}
\put(736,411){\rule[-0.175pt]{0.350pt}{9.154pt}}
\put(726,411){\rule[-0.175pt]{4.818pt}{0.350pt}}
\put(726,449){\rule[-0.175pt]{4.818pt}{0.350pt}}
\put(861,312){\rule[-0.175pt]{0.350pt}{8.191pt}}
\put(851,312){\rule[-0.175pt]{4.818pt}{0.350pt}}
\put(851,346){\rule[-0.175pt]{4.818pt}{0.350pt}}
\put(897,282){\rule[-0.175pt]{0.350pt}{14.454pt}}
\put(887,282){\rule[-0.175pt]{4.818pt}{0.350pt}}
\put(887,342){\rule[-0.175pt]{4.818pt}{0.350pt}}
\put(1025,228){\rule[-0.175pt]{0.350pt}{6.745pt}}
\put(1015,228){\rule[-0.175pt]{4.818pt}{0.350pt}}
\put(1015,256){\rule[-0.175pt]{4.818pt}{0.350pt}}
\put(1054,221){\rule[-0.175pt]{0.350pt}{8.191pt}}
\put(1044,221){\rule[-0.175pt]{4.818pt}{0.350pt}}
\put(1044,255){\rule[-0.175pt]{4.818pt}{0.350pt}}
\put(781,370){\rule[-0.175pt]{0.350pt}{8.913pt}}
\put(771,370){\rule[-0.175pt]{4.818pt}{0.350pt}}
\put(771,407){\rule[-0.175pt]{4.818pt}{0.350pt}}
\put(897,292){\rule[-0.175pt]{0.350pt}{5.782pt}}
\put(887,292){\rule[-0.175pt]{4.818pt}{0.350pt}}
\put(887,316){\rule[-0.175pt]{4.818pt}{0.350pt}}
\sbox{\plotpoint}{\rule[-0.250pt]{0.500pt}{0.500pt}}%
\put(1006,632){\makebox(0,0)[r]{quark model}}
\put(1028,632){\usebox{\plotpoint}}
\put(1048,632){\usebox{\plotpoint}}
\put(1069,632){\usebox{\plotpoint}}
\put(1090,632){\usebox{\plotpoint}}
\put(1094,632){\usebox{\plotpoint}}
\put(264,158){\usebox{\plotpoint}}
\put(264,158){\usebox{\plotpoint}}
\put(271,177){\usebox{\plotpoint}}
\put(279,196){\usebox{\plotpoint}}
\put(287,215){\usebox{\plotpoint}}
\put(295,234){\usebox{\plotpoint}}
\put(303,253){\usebox{\plotpoint}}
\put(311,273){\usebox{\plotpoint}}
\put(319,292){\usebox{\plotpoint}}
\put(328,311){\usebox{\plotpoint}}
\put(337,329){\usebox{\plotpoint}}
\put(346,348){\usebox{\plotpoint}}
\put(355,366){\usebox{\plotpoint}}
\put(366,385){\usebox{\plotpoint}}
\put(376,403){\usebox{\plotpoint}}
\put(386,421){\usebox{\plotpoint}}
\put(398,438){\usebox{\plotpoint}}
\put(410,455){\usebox{\plotpoint}}
\put(422,471){\usebox{\plotpoint}}
\put(436,487){\usebox{\plotpoint}}
\put(450,501){\usebox{\plotpoint}}
\put(467,514){\usebox{\plotpoint}}
\put(485,525){\usebox{\plotpoint}}
\put(503,534){\usebox{\plotpoint}}
\put(524,538){\usebox{\plotpoint}}
\put(544,540){\usebox{\plotpoint}}
\put(565,538){\usebox{\plotpoint}}
\put(585,533){\usebox{\plotpoint}}
\put(604,526){\usebox{\plotpoint}}
\put(623,517){\usebox{\plotpoint}}
\put(641,506){\usebox{\plotpoint}}
\put(659,496){\usebox{\plotpoint}}
\put(675,483){\usebox{\plotpoint}}
\put(692,471){\usebox{\plotpoint}}
\put(708,458){\usebox{\plotpoint}}
\put(724,445){\usebox{\plotpoint}}
\put(740,432){\usebox{\plotpoint}}
\put(756,418){\usebox{\plotpoint}}
\put(772,405){\usebox{\plotpoint}}
\put(788,392){\usebox{\plotpoint}}
\put(804,378){\usebox{\plotpoint}}
\put(820,365){\usebox{\plotpoint}}
\put(836,352){\usebox{\plotpoint}}
\put(853,340){\usebox{\plotpoint}}
\put(869,327){\usebox{\plotpoint}}
\put(886,315){\usebox{\plotpoint}}
\put(903,303){\usebox{\plotpoint}}
\put(920,291){\usebox{\plotpoint}}
\put(937,280){\usebox{\plotpoint}}
\put(955,269){\usebox{\plotpoint}}
\put(973,259){\usebox{\plotpoint}}
\put(992,249){\usebox{\plotpoint}}
\put(1011,241){\usebox{\plotpoint}}
\put(1030,232){\usebox{\plotpoint}}
\put(1049,225){\usebox{\plotpoint}}
\put(1068,218){\usebox{\plotpoint}}
\put(1088,210){\usebox{\plotpoint}}
\put(1108,205){\usebox{\plotpoint}}
\put(1128,199){\usebox{\plotpoint}}
\put(1136,197){\usebox{\plotpoint}}
\end{picture}
\end{centering}
\caption{The radial wavefunctions for the $1S$ and $1P$ charmonium
computed using improved actions and two different lattice spacings.
Wavefunctions from a continuum quark model are also shown. Statistical
errors are negligible for the $1S$~wavefunction.}
\label{wfcns}
\end{figure}
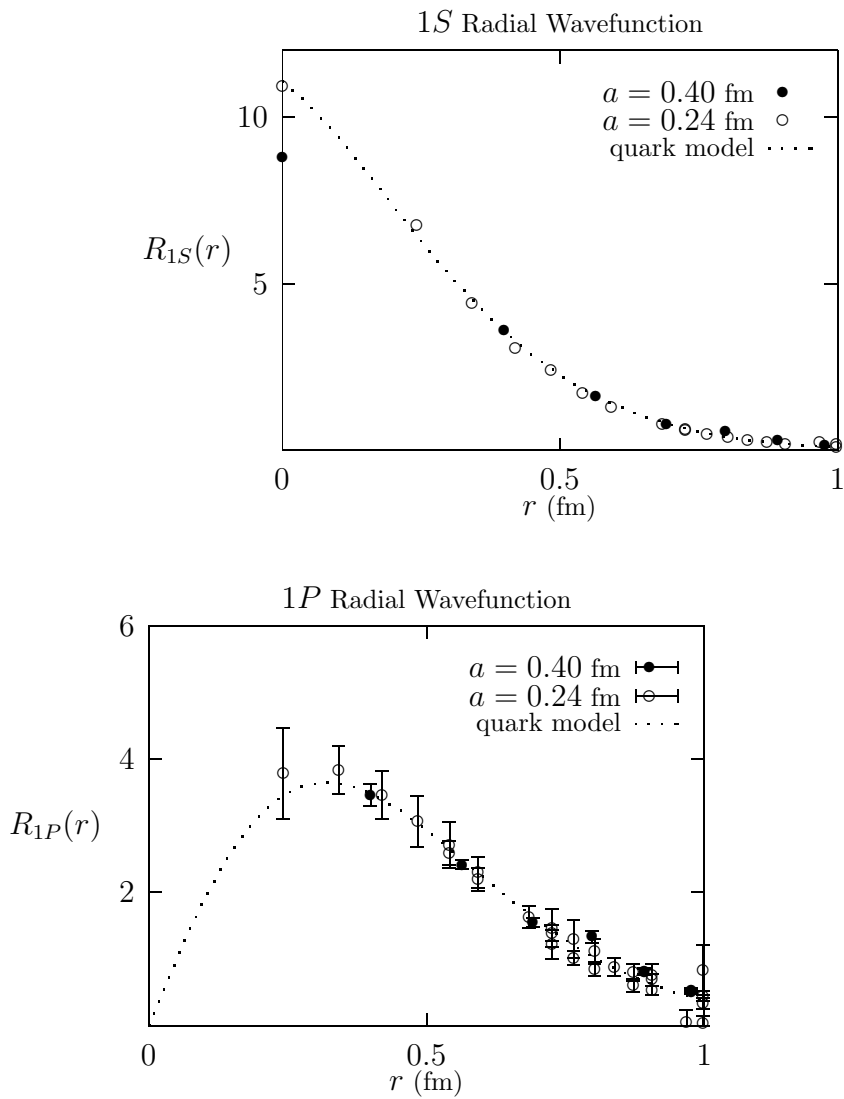

In Fig.~\ref{wfcns} we show simulation results for
the Coulomb-gauge radial wavefunctions of $1S$ and $1P$ charmonium as
computed with the improved action on our coarsest and finest lattices.
We also show wavefunctions computed from a continuum quark
model tuned to reproduce our lattice results.
The wavefunctions from the different lattice spacings agree
well everywhere except at $r=0$, where agreement is not expected
(because of renormalization effects). The wavefunctions show
remarkable rotational symmetry considering that the rms radius of the
$1S$~wavefunction, for example, is just 0.37~fm, or slightly less than one
lattice spacing on the coarse lattice\,---\,accurate modeling of a
hadron is possible with improved actions
even when the hadron is only a few lattice spacings in extent.

We also calculated the ratio of the $1P$-$1S$~charmonium
splitting to the square
root of the slope~$V^\prime$ of the static potential at
$r=0.6$~fm\,\cite{som}; our results are in Table~\ref{scaling}.
(We choose 0.6~fm because we have good simulation results for $V(r)$ at that
radius and, also, $V^\prime$ is almost independent of $r$ there.)
This ratio is a dimensionless quantity that should
become independent of lattice spacing as we approach the
continuum limit. We see that, to within our errors, the improved action
has reached the continuum limit at a lattice spacing of $0.4$~fm,
whereas the Wilson action has not. The twisted-rectangle action also
gives excellent results at this lattice spacing.

\begin{table}[t]
\begin{center}
\begin{tabular}{lcccc} \hline \\
Action & $\beta_\pl$ & $a (1P-1S)$ & $a$ & $(1P-1S)/\sqrt{V^\prime}$ \\ \hline
Wilson & 4.5 & 1.410 (30) & .41~fm & 1.38 (5) \\
& 5.7 & \dgsp .383 (10) & .17~fm & \dgsp .87 (3) \\ \mbox{} \\
Improved & 6.8 & \dgsp .922 (\dgsp 6) & .40~fm & \dgsp .90 (1) \\
& 7.1 & \dgsp .756 (\dgsp 4) & .33~fm & \dgsp .92 (1) \\
& 7.4 & \dgsp .558 (\dgsp 6) & .24~fm & \dgsp .89 (2) \\ \mbox{} \\
$S_\trt$ & 4.1 & \dgsp .930 (30) & .40~fm &  \dgsp .92 (3) \\ \hline \hline
\end{tabular}
\end{center}
\caption{Ratio of the charmonium $1P-1S$~splitting to the square root of the
derivative of the static potential $V(r)$ at 0.6~fm. Charmonium results for
the Wilson action at $\beta=5.7$ are from~[6].}
\label{scaling}
\end{table}

Since coarse lattices have far fewer sites and much less
critical-slowing-down, the cost to produce a statistically independent
configuration should be much less on a coarser lattice. To examine
this issue, we compared our results for $V(r)$
with those in~\cite{bs} which are for the
potential computed using the Wilson action at $\beta=6$ ($32^4$
lattice with $a\approx0.1$~fm). We rescaled the coordinates and
potential from this other study to put them in the same units as our
$\beta_\pl=6.8$ results, and examined the potentials at comparable
distances. The results from both simulations are listed in
Table~\ref{bspotl}. In both cases the potential is obtained from the
time dependence of loop-like correlation functions for times equal to
or larger than some~$T_{\rm min}$. Results should become independent
of $T_{\rm min}$ once it is sufficiently large; with
our much smaller statistical
errors, we can measure the small shift when $T_{\rm min}$ is doubled from
.4~fm to .8~fm. Our results required
$1.3\times10^7$ site updates, while the analysis on the fine lattice
required $6.4\times10^9$ site updates. Since statistical errors (for
$T_{\rm min}=0.4$~fm) are about 20~times smaller for the coarse
lattice, we estimate that comparable errors with the fine lattice
would require 197,000~times more site updates than we used on the
coarse lattice.

\begin{table}[thb]
\begin{center}
\begin{tabular}{clll} \hline \\
 $r/a_c$ & \multicolumn{2}{c}{$a=a_c= .40$} &
   \multicolumn{1}{c}{$a= .10$} \\
 & $T_{\rm min}=.4$ & $T_{\rm min}=.8$ & $T_{\rm min}=.3$ \\ \hline
 \dgsp .962 &&& \dgsp .839 (11) \\
 1 & \dgsp .871 (\dgsp 0) & \dgsp .887 (\dgsp 1) & \\
 \mbox{} && \\
 1.361 &&& 1.316 (25) \\
 $\sqrt{2}$ & 1.373 (\dgsp 1) & 1.384 (\dgsp 2) & \\
 \mbox{} && \\
 1.667 &&& 1.581 (54) \\
 $\sqrt{3}$ & 1.718 (\dgsp 2) & 1.742 (10) & \\
 \mbox{} && \\
 1.924 &&& 1.842 (42) \\
 2 & 1.897 (\dgsp 2) & 1.941 (10) & \\ \hline \hline
 \end{tabular}
 \caption{Comparison of the static-quark potential $a\,V(r)$  as computed on
a coarse lattice (improved action, $\beta_\pl=6.8$,
$a= 0.40$~fm), and on a fine lattice (Wilson action, $\beta=6$,
$a= 0.10$~fm). $T_{\rm min}$ is the shortest time interval used
in the correlation functions that determine $V(r)$. All times and
distances are in fm.}
\label{bspotl}
\end{center}
 \end{table}

\section{Conclusions}

We have found that, by using  perturbatively-improved  actions with
tadpole-improved operators, we can
accurately simulate quark and gluon dynamics on lattices with a spacing as
large as 0.4~fm.  Using either of two very different improved gluon
actions, we obtained results for the confining potential and for
charmonium that are independent of finite-lattice-spacing artifacts to within
a few percent.

It is striking that we can obtain accurate results
on such coarse lattices using relatively simple actions.
Following the ideas of the Wilson renormalization group,
one expects, even at fixed, finite lattice spacing,
to achieve arbitrarily accurate results by adding more and more operators
to the discretized action, and by calculating their coefficients with
increasing accuracy.
We find that including just the leading corrections, with coefficients
calculated in (tadpole-improved) tree-level perturbation theory, gives
excellent results even on our coarsest lattices.
Ultimately, nonperturbative calculations (for example, using Monte
Carlo renormalization group methods) should be used to determine the
coefficients, at least as
a check of the perturbation theory.
Such nonperturbative determinations of the couplings in the lattice action
would be very interesting, although we expect such an effort to be
much more difficult than the tadpole-improvement program outlined
in this paper, and to produce only
small changes to the coefficients determined here.  In~\cite{lm}, the
normalization of the strong coupling constant and the leading gluonic
operator ($F_{\mu\nu}^2$) was carefully studied.  We found that
tadpole-improved perturbation theory gave
normalizations that were extremely close
to the nonperturbatively obtained normalization.

The coarseness of the lattice
makes our simulations $10^4$--$10^5$ times faster than ones using
unimproved actions; in fact, most of the $a=0.4$~fm
results in this paper were
obtained using an IBM RS6000/250 desktop workstation,
which is powered by a personal-computer CPU
(66MHz PowerPC).  We therefore believe that, using the methods we
have described, the QCD hadron spectrum can be calculated to an
accuracy of a few percent using computer resources that are already
widely available. And by combining these techniques
with forefront computing technology, we can
begin to tackle problems in nonperturbative QCD far more complex than
previously imagined.

\centerline{\bf Acknowledgments}
\smallskip
We are very grateful to A.~Pierce, IBM, and Cornell's Center for
Theory and Simulation
for help in using their IBM SP-2 supercomputer for our larger
calculations.
Some calculations in this work were performed on the Fermilab lattice
supercomputer, ACPMAPS.
This work was supported by the DOE and NSF.

\end{document}